\newcommand{\doublespace}{
  \renewcommand{\baselinestretch}{1.5}
  \large\normalsize}
\documentstyle[preprint,aps,tighten,prb]{revtex}
\begin{document}
\doublespace
\title{Systematic study of Ga$_{1-x}$In$_x$As self-assembled
quantum wires with different interfacial strain relaxation}

\author{Liang-Xin Li, Sophia Sun and Yia-Chung Chang}
\address{Department of Physics and Materials Research Laboratory 
\\ University of Illinois at Urbana-Champaign, Urbana, Illinois 61801}

\date{\today}
\maketitle

\begin{abstract}
A systematic theoretical study of the electronic and optical properties of
Ga$_{1-x}$In$_x$As self-assembled quantum-wires (QWR's) made of 
short-period superlattices (SPS) with strain-induced lateral ordering 
is presented.
The theory is based on the effective bond-orbital model (EBOM) combined with a valence-force field (VFF)
model. Valence-band anisotropy, band mixing, and effects due to 
local strain distribution at the atomistic level are all taken into account.
Several structure models with varying degrees of alloy mixing for lateral modulation are considered. 
A valence force field model is used to find the equilibrium atomic
positions in the QWR structure by minimizing the lattice energy.
The strain tensor at each atomic (In or Ga) site is then obtained
and included in the calculation of electronic states and optical properties. 
It is found that different local arrangement of atoms leads to very different
strain distribution, which in turn alters the optical properties.
In particular, we found
that in model structures with thick capping layer the electron and hole are 
confined in the Ga-rich region and the optical anisotropy can be reversed due 
to the variation of lateral alloying mixing, while for model structures 
with thin capping layer the 
electron and hole are confined in the In-rich region, and the optical anisotropy
is much less sensitive to the lateral alloy mixing. 
\end{abstract}

\mbox{}

\mbox{}

\newpage

\section{Introduction}

Optical properties of III-V semiconductor nanostructures have attracted 
a great deal of interest recently for their applications in optical communications that 
involve switching, amplification, and signal processings.
$Ga_xIn_{1-x}As$ and $Ga_xIn_{1-x}P$ 
are among   the most important ternary III-V 
compound semiconductors. The band gap of Ga$_x$In$_{1-x}$As covers
both the 1.3 and 1.55 $\mu m$ wavelengths, which are preferred in long distance fiber communications. However, long-wavelength photonic devices based on 
lattice-matched Ga$_{0.47}$In$_{0,53}$As/InP heterostructures suffer 
from strong Auger recombination and inter-valence band absorption 
processes[1-3].
Recently, to improve the performance of long-wavelength semiconductor 
lasers, long wavelength($\sim 1.55 {\mu}$m) Ga$_x$In$_{1-x}$As quantum-wire 
(QWR) lasers have been grown by a single step molecular beam 
epitaxy technique[4-6].

An important optical property is the change in the emission of light 
(in energy, polarization, 
and intensity) that results from phase-space filling of carriers in 
one- and two-dimensionally (2D) confined 
systems. As the dimensionality of the quantum confinement increases 
from 1D to 2D, 
the sharpening of the density of states gives rise to a lower excitation 
threshold, thereby yielding potentially enhanced optical effects.
It is found that the QWR laser structure is a
promising candidate for optical communication device because of 
the many predicted benefits, such as higher gain, 
reduced temperature sensitivity, higher modulation bandwidths, and narrower 
spectral linewidths[5]. 

The fabrication of quantum wires via the strain-induced lateral-layer ordering
(SILO) process starts with the growth of
short-period superlattices (SPS) [e.g. (GaAs)$_2$/(InAs)$_{2.25}$]
along the [001] direction. The excess fractional InAs layer leads to
stripe-like islands during the MBE growth.[4] The presence of stripes combined with the strain effect lead to a 
natural phase separation as
additional layers of GaAs or InAs are deposited.
A self-assembled quantum wire heterostructure can then be
created by sandwiching the composition modulated layer
between barrier materials such as Al$_{0.24}$Ga$_{0.24}$In$_{0.52}$As (quarternary), Al$_{0.48}$In$_{0.52}$As(ternary), or InP (binary)[4-6].
It was found that different barrier materials can lead to different degree of
lateral composition modulation, and the consequent optical properties are different.[6] Besides the self-assembled lateral
ordering, it is believed that the strain also plays a key role[5,6] 
in the temperature stability and optical anisotropy for the QWR laser structure.
Much work has been undertaken which 
predicts that by using strained-layer superlattice to form the active 
region of a 
quantum-wire laser, the threshold current can be decreased by one order of 
magnitude, and the optical loss due to intervalence-band absorption and  Auger 
recombination will also be greatly reduced[4-6]. Also, the temperature 
sensitivity 
is reduced by an order of magnitude compared with strain-free structures.
Temperature sensitivity of the lasing wavelength for typical GaInAs/InP lasers
is around $ 5 \AA/{}^0C$. By using a distributed-feedback structure, the temperature dependence of the lasing wavelength can be reduced to $1\AA/{}^0C$[5]. With the self-assembled 
GaInAs QWR, the temperature sensitivity is smaller than $0.1\AA/{}^0C$[5,6]. 
Thus, it represents an important improvement on the current technology in fabricating 
long wavelength lasers for fiber communication. 

	A few theoretical studies of Ga$_x$In$_{1-x}$As self-assembled QWRs 
based on a simplified uniform strain model have been reported.[8-10] 
In these calculations the SPS region is modeled by a Ga$_x$In$_{1-x}$As alloy 
with a lateral modulation of the composition $x$. Although these calculations can explain
the QWR band gap and optical anisotropy qualitatively, it does not take into 
account the detailed SPS structure and the microscopic strain distribution.
The understanding of these effects is important if one wish to have a full 
design capability of the self-assembled QWR optoelectronic devices. 
On the other hand, microscopic strain distributions in several SPS structures 
have been
calculated, and the electronic properties of these SPS structures have been 
studied via an empirical pseudopotential method[11]. 
However, the effects of sandwiching the SPS structures between barrier 
materials so as to form QWRs  have not been
explored. It is also worth pointing out that  
the effective masses obtained in the pseudopotential method for the 
constituent bulk materials are 0.032$m_0$ and 0.092$m_0$ 
for InAs and GaAs, respectively.[12] These values are  30\% larger
than the actual values (0.024$m_0$ and 0.067$m_0$); 
thus, the energy levels of quantum confined states obtained by this method 
are subject to similar uncertainty.  

In this paper we present a systematic study of the effects of microscopic 
strain distribution on
the electronic and optical properties of the Ga$_{1-x}$In$_x$As 
self-assembled QWRs via the combination of
the effective bond-orbital model (EBOM) for electronic states and the 
valence-force-field (VFF) model for microscopic strain distribution.
Both clamped and unclamped SPS structures with different degrees of lateral 
alloy mixing are considered. The clamped structure
has a SPS layer sandwiched between the substrate and a thick capping layer
so that the interface between the top monolayer of the SPS and the barrier 
region is atomically flat.
The unclamped structures correspond to a SPS layer sandwiched between the 
substrate and a thin capping layer, which allows more flexible relaxation of
interface atoms so the strain energy can be relieved. 
This leads to a wavy interface structure. Most self-assembled QWR 
samples reported to date are closer to unclamped structures with a wavy 
interface, although it should be possible to produce self-assembled QWRs 
closer to the clamped structures by using a thick capping layer.

Our theoretical studies show that there are profound differences in the 
electronic and optical properties between the clamped and unclamped self-assembled QWR 
structures. In particular, we found
that in clamped structures the electron and hole are confined in the Ga-rich 
region and the polarization of Photoluminescence (PL) can go from 
predominantly along [110] for SPS with 
abrupt change in In/Ga composition to predominantly along [$1\bar 10$] for 
SPS with smooth change in In/Ga composition. On the other hand, for 
unclamped structures the 
electron and hole are confined in the In-rich region, and the optical 
polarization is always predominately along [$1\bar 10$] with a 
weak dependence on the lateral alloy mixing. 
This also implies that by changing the degree of strain relaxation at the 
interface between the  SPS and the capping layer, one can tailor the optical 
properties of self-assembled QWRs. Thus, through this theoretical study, 
we can gain a better understanding of the strain engineering of  
self-assembled QWR structures which may find applications in fiber-optical 
communication. 

\section {Model structures}

For both clamped and unclamped SPS structures, we consider different degrees of
lateral alloy mixing and examine the effects of
the microscopic strain distribution on the electronic and optical properties.
Two example QWR model structures 
considered in the present paper (prior to alloy mixing) are depicted
in Figure 1. The supercell of the first model structure consists
of 8 pairs of (001) (GaAs)$_2$/(InAs)$_2$ SPS
with a total thickness of $\approx 94 \AA$ (quantum well region)
followed by a Al$_{0.24}$Ga$_{0.24}$In$_{0.52}$As layer (barrier region)
with thickness $\approx 60 \AA$ (20 diatomic layers).
The supercell of the second model structure consists
of 8 pairs of (001) (GaAs)$_2$/(InAs)$_{2.25}$ SPS
with a total thickness of $\approx 100 \AA$ (quantum well region)
followed by a Al$_{0.24}$Ga$_{0.24}$In$_{0.52}$As layer ( barrier region).
We assume an arrangement of alternating stripe like islands due to strain 
induced lateral ordering. In the diagram, no lateral alloy mixing is shown.
In our calculations of strain distribution and electronic structures, different
degrees of lateral alloy mixing will be considered.
Experimentally, self-assembled 
InGaAs QWRs were usually grown with the (2/2.25) SPS structure.[12] 
However, the migration
of excess In atoms during MBE growth could lead a structure somewhere between (2/2) SPS and (2/2.25) SPS structures.

In both model structures, we can divide the self-assembled QWR into
two regions with the left half being Ga rich and the right half In rich.
During growth, varying degrees of lateral alloy
mixing of these islands with the surround atoms is likely to occur.
In the atomic layers with In-rich alloy filled the right half of the
unit cell (layer 2 in structure 1 and layers 7 and 9 in structure 2),  
the In composition $x_{In}$ is 
assumed to vary in the [110] direction (or $y'$ direction) 
according to the relation
\begin{equation}
x_{In}=  \left\{ \begin{array}{ll}
 x_m [1-\sin (\pi y'/2b)] /2 &  \mbox{ for } y'< b\\ 
	0  & \mbox{ for } b < y' < L/2-b \\
 x_m \{1 + \sin [\pi (y'-L/2)/2b]\} /2 &  \mbox{ for } L/2-b < y' < L/2+b \\
 x_m   &	  \mbox{ for } L/2+b < y' < L-b \\
 x_m \{1 - \sin [\pi (y'-L)/2b]\} /2 &  \mbox{ for } y'> L-b, \end{array} \right.
\end{equation}
where $x_m$ is the maximum In composition in the layer, $2b$ denotes the 
width of lateral composition grading, and $L$ is the period of the
lateral modulation in the [110] direction.
In layer 4 of structure 1, which contains Ga-rich alloy in the left half of the
unit cell, the Ga composition ($x_{Ga}$) as a function of $y'$ is given by
a similar equation with the sign of the sine function reversed as compared to
Eq. (1). In structure 2, there are a few atomic layers
that contain 0.25 monolayer of In (layers 3 and 13) or Ga( layers 5 and 11).
The lateral alloy modulation in layers 3 and 13 is described by
\begin{equation}
x_{In}=  \left\{ \begin{array}{ll}
 0   &    \mbox{ for } 0 < y' < 5L/8-b \\
 x_m \{1 + \sin [\pi (y'-5L/8)/2b]\} /2 &  \mbox{ for } 5L/8-b<y'< 5L/8+b,\\ 
 	x_m  & \mbox{ for } 5L/8+b < y' < 7L/8-b \\
 x_m \{1 - \sin [\pi (y'-7L/8)/2b]\} /2 &  \mbox{ for } 7L/8-b < y' < 7L/8+b \\
 0   &    \mbox{ for } 7L/8+b < y' < L.
\end{array} \right.
\end{equation}
Similar equation for $x_{Ga}$ in layers 5 and 11 can be deduced
from the above.
By varying the parameters $x_m$ and $b$, we can get different degrees
of lateral alloy mixing. Typically $x_m$ is between 0.6 and 1, and $b$
is between zero and $15 a_{[110]} \approx 62 \AA$.

A valence force field (VFF) model[13-15] is used to find the equilibrium 
atomic positions in the self-assembled QWR structure by minimizing the lattice energy.
The strain tensor at each atomic (In or Ga) site is then obtained by
calculating the local distortion of chemical bonds.
We found that different local arrangement of atoms can lead to very different
strain distribution. In particular, the shear strain in the clamped structure
can change substantially when the In/Ga composition modulation is changed.
Consequently, the optical anisotropy can be reversed due to the change in 
the strength of the shear strain caused by
the intermixing of In and Ga atoms.

\section{Theoretical Approach}
The method used in this paper for calculating the strained QWR band 
structure is based 
on the effective bond-orbital model (EBOM). A detailed description of this 
method has been published elsewhere[16,17].
EBOM is a tight binding-like model with minimum set of localized basis
(the bonding or anti-bonding orbitals).
The interaction and optical parameters are obtained by a correspondence
with the ${\bf k\cdot p}$ theory, which can be cast into analytic forms. 
Thus, the model can be viewed as a spatially
discretized version of the ${\bf k\cdot p}$ method, while retaining the 
virtues of LCAO (linear combination of atomic orbitals) method.
The ${\bf k\cdot p}$ model is the most popular one for treating electronic
structures of semiconductor quantum wells or superlattices.
However, when applied to complex structures such as self-assembled 
quantum wires[4-9] or quantum dots[13,18,19], the method becomes very cumbersome
if one wish to implement the correct boundary conditions that
take into account the differences in
${\bf k\cdot p}$ band parameters for different materials involved.
EBOM is free of this problem, since different material parameters are
used at different atomic sites in a natural way. For simple structures, when
both EBOM and ${\bf k\cdot p}$ model are equally applicable,
the results obtained are essentially identical.[16]

The optical matrix elements for the QWR states are computed
in terms of elementary optical matrix elements between the valence-band
bond orbitals and the conduction-band orbitals. 
The present calculation includes the coupling of the four spin-3/2 
valence bands and the two spin-1/2 conduction bands closest to the band edges.
Thus, it is equivalent to a 6-band ${\bf k\cdot p}$ model.
For our systems studied here, the band-edge properties are relatively 
unaffected by the split-off band due to the large spin-orbit splitting
as discussed in our previous paper[9]. Hence, the split-off bands are 
ignored here. The bond-orbitals for the GaAs and InAs	needed in
the expansion of the superlattice states contain the following: 
four valence-band bond orbitals per bulk unit cell, which are p-like 
orbitals coupled with the spin to form orbitals with total 
angular momentum J=3/2 plus two conduction-band bond orbitals with J=1/2. 
They are written as
\begin{equation}
|{\bf R},u_{JM}>=\sum_{\alpha,\sigma}C(\alpha,\sigma,J,M)|\vec{R},\alpha>
\psi_{\sigma},
\end{equation}
where $J=1/2$ and 3/2 for the conduction and valence bands, respectively, 
and $M=-J,\cdots J$, $\psi_{\sigma}$, designates the electron 
spinor($\sigma$=1/2,-1/2), and $|{\bf R},\alpha>$ denote an 
$\alpha$-like($\alpha=s,x,y,z$) bond orbital, located at unit cell 
${\bf R}$. $C(\alpha,\sigma,J,M)$ are the coupling coefficients 
obtainable by group theory. All these bond orbitals are assumed 
to be sufficiently localized so that the interaction between orbitals 
separated farther than the nearest-neighbor distance can be ignored. 

    The effect of strain is included by adding a strain Hamiltonian 
$H^{st}$ to the EBOM Hamiltonian[17].  The matrix elements 
of $H^{st}$ in the bond-orbital basis can be 
obtained by the deformation-potential theory of Bir and Pikus[20]. 
We use the valence-force field (VFF) model of Keating[14] and Martin[15]
to calculate the microscopic strain distribution. This model has been shown 
to be successful in fitting and predicting the elastic constants of 
elastic continuum theory, calculating strain distribution in a quantum well, 
and determining the atomic structure of III-V alloys. 
It was also used in the calculation of the strain distribution in 
self-assembled quantum dots.[13,19] The VFF model is a microscopic theory 
which includes bond stretching and bond bending, and avoids the potential 
failure of elastic continuum theory in the atomically thin limit. 
The total energy of the lattice is taken as 

\begin{equation}
V =\frac{1}{4}\sum_{ij}\frac{3}{4}\alpha_{ij}(d^2_{ij}-d^2_{0.ij})^2/d^2_{0,ij}
+\frac{1}{4}\sum_i\sum_{j\neq k}\frac{3}{4}\beta_{ijk}(\vec{d_{ij}}.\vec{d_{ik}}
+d_{0,ij}d_{0,ik}/3)^2/d_{0,ij} d_{0,ik}
\end{equation}

\noindent where $i$ runs over all the atomic sites, $j, k$ run over the 
nearest-neighbor sites of $i$, $\vec{d_{ij}}$ is the vector joining the 
sites i and j, $d_{ij}$ is the length of the bond, $d_{0,ij}$ is the 
corresponding equilibrium length in the 
binary constituents, and $\alpha_{ij}$ and $\beta_{ijk}$ are the 
bond stretching and bond bending constants, respectively. 
$\alpha$ and $\beta$ are from Martin's calculations.[11] 
For the bond-bending
parameter $\beta$ of In-As-Ga, we take $\beta_{ijk}=\sqrt{\beta[ij]\beta[ik]}
$ following Ref. [15]. 
     
  To find the strain tensor in the $InAs/GaAs$ self-assembled QWR, we start from  
ideal atomic positions and minimize the system energy 
using the 
Hamiltonian given above. Minimization of the total energy requires one to 
solve a set of coupled equations with 3N variables, where N is the total 
number of atoms. Direct solution of these equations is impractical 
in our case, since the system contains more than 6,000 atoms. 
We use an approach taken by several authors[10,13,19] which has been shown 
to be quite efficient. In the beginning of the simulation all the atoms are 
placed on the InP lattice, we allow atoms to deviate from this starting 
positions and use periodic boundary conditions in the plane
perpendicular to the growth direction, while keeping atoms in the
planes outside the SPS region at their ideal atomic positions
for a InP lattice (since the self-assembled QWR is grown epitaxially on the InP 
substrate). In each iteration, only one atomic 
position is displaced and other atom positions are held fixed. 
The direction of the displacement of atom $i$ is determined according to
the force  $f_i=-\partial V/\partial x_i$ acting on it.
All atoms are displaced in sequence. the whole sequence is
repeated until the forces acting on all atoms become zero
at which point the system energy is a minimum.
Once the positions of all the atoms are known, the strain distribution 
is obtained through the strain tensor calculated according to the method
described in Ref. [21].

Let $R^0$ be the position matrix without strain:
\[
R^0  = \left ( \begin{array}{ccc}
       R^0_{12x} & R^0_{23x} & R^0_{34x} \\
       R^0_{12y} & R^0_{23y} & R^0_{34y} \\
       R^0_{12z} & R^0_{23z} & R^0_{34z}  
      \end{array} \right ), \]
where ${\bf R}^0_{ij} = {\bf R}^0_j - {\bf R}^0_i; i,j=1,4$ and ${\bf R}^0_i$
$(i=1,4)$ denote positions of four As atoms surrounding a Ga or In atom.
Here we choose ${\bf R}^0_{12} = (1, -1, 0)a/2$, 
${\bf R}^0_{23} = (-1, 0, -1)a/2$, and ${\bf R}^0_{34} = (1, 1, 0)a/2$. 
$a$ is the lattice constant of GaAs or InAs, depending on the site.
Let $R$ be the corresponding position matrix with strain. It was shown
in Ref. [21] that the strain tensor 
is given by

\begin{equation}
\epsilon=R*R_0^{-1} - 1.
\end{equation}

After getting the strain tensor, the strain Hamiltonian is given by Bir and
Pikus[20]

\begin{equation}
H^{st} = \left( \begin{array}{ccc} 
 -\Delta V_H + D_1 &\sqrt{3}d e_{xy} &\sqrt{3}d e_{xz} \\
  \sqrt{3}d e_{xy} & -\Delta V_H + D_2     &\sqrt{3}d e_{yz} \\ 
       \sqrt{3}d e_{xz} &\sqrt{3}d e_{yz} & -\Delta V_H + D_3
               \end{array} \right),
\end{equation}
where $e_{ij}=(\epsilon_{ij}+\epsilon_{ji})/2$, and
$$
\Delta V_H = (a_1+a_2)(\epsilon_{xx}+\epsilon_{yy}+\epsilon_{zz}),~~\\
 D_1 = b(2\epsilon_{xx}-\epsilon_{yy}-\epsilon_{zz}),~~\\
 D_2 = b(2\epsilon_{yy}-\epsilon_{xx}-\epsilon_{zz}),~~\\
 D_3 = b(2\epsilon_{zz}-\epsilon_{yy}-\epsilon_{xx}),\\
$$
The strain potential on the s states is given by
$$\Delta V_c = c_1(\epsilon_{xx}+\epsilon_{yy}+\epsilon_{zz}),$$
The strain Hamiltonian in the bond-orbital basis $|JM>$ can be easily 
found by using the coupling constants[13], i.e, 
\begin{equation}
<JM|H^{st}|J'M'>=\sum_{\alpha, \alpha',\sigma}C(\alpha, \sigma;J,M)^*
C(\alpha', \sigma;J',M')H^{st}_{\alpha\alpha'}
\end{equation}
The elastic constants $C_{12}$ and $C_{11}$ for GaAs, InAs and AlAs can be found
in Ref. [22,23]. The deformation potentials $a_1,~a_2,~b,~c_1, ~d$ can be
found in Ref.[23-26]. The linear interpolation
and virtual crystal approximations used to obtain the corresponding
parameters for the barrier material (Al$_{0.24}$Ga$_{0.24}$In$_{0.52}$As).

The above strain Hamiltonian is derived locally for the each cation atom in 
the self-assembled QWR considered. To calculate the electronic states of the 
self-assembled QWR for model structure 1[see Fig. 1(a)],
we first construct a zeroth-order
Hamiltonian for a superlattice structure which contains in each period
8 pairs of (001) (GaAs)$_{2}$/(InAs)$_{2}$ 
SPS layers (with a total thickness around 100 \AA) and 20 diatomic layers of 
Al$_{0.24}$Ga$_{0.24}$In$_{0.52}$As (with thickness around 60 \AA).
So, the superlattice unit cell for the zeroth-order model contains
52 diatomic layers.  The appropriate strain Hamiltonian for the
the (GaAs)$_{2}$/(InAs)$_{2}$ SPS  on InP is also included.
For model structure 2 [see Fig. 1(b)], the zero-order superlattice
consistes of two additional monolayers of InAs 
inserted into the 8 pairs of (2/2) SPS:
one between the 2nd and 3rd pair of (2/2) SPS, the other between the 
5th and 6th pair of (2/2) SPS. 

The eigen-states for the zero-th order Hamiltonian for different
values of $k_2$ (separated by the SL reciprocal lattice vectors
in the [110] direction) are then used as the basis
for calculating the self-assembled QWR electronic states.
The difference in the Hamiltonian (including strain effects)
caused by the intermixing of Ga and In
atoms at the interfaces is then added to the zeroth-order Hamiltonian,
and the electronic states of the  full Hamiltonian is solved by 
diagonalizing the Hamiltonian matrix defined within a truncated set of
eigen-states of the zeroth-order Hamiltonian. A total of $\sim 500$ eigen-states
of the zero-th order Hamiltonian (with 21 different $k_{110}$ points)
were used in the expansion. The subbands
closest to the band edge are converged to within 0.1 meV.

\section{Results and discussions}

{\bf A. Strain distributions}

In this section we discuss strain distributions in our model structures
as described in section I with varying degree of lateral alloy mixing.
Both clamped (with thick capping layer) and unclamped (with thin capping layer)
situations are considered.
To model the alloy structure with composition modulation,
we use a super-cell
which contains 72 atoms in the [110] ($y'$) direction, 36 atoms in the
[$1\bar 1 0$] ($x'$) direction and 64 or 68 atomic planes along the [001] ($z$)
direction [8 pairs of (2/2) or (2/2.25) SPS] plus a GaAs capping layer. 
In the atomic planes which consist of alloy structure, we first determine
the In composition at a given $y'$ according to Eq. (1) and then use a
random number generator to determine the atomic species along the
$x'$ direction. The bottom layer of atoms are bonded to the InP substrate
with the substrate atoms fixed at their ideal atomic positions.
The calculated strain distributions are then averaged over
the $x'$ coordinate. 
For the clamped case,  the GaAs capping layer is assumed to be lattice
matched to InP with a flat surface. For the unclamped case, the capping
layer is allowed to relax freely, thus giving rise to a wavy surface structure.

The diagonal strains in
the four atomic layers that constitute the (2/2) SPS in structure 1 for 
the clamped and unclamped cases are shown in Figs. 2 and 3, respectively.
The lateral alloy modulation considered is described by Eq. (1)
with $x_m=0.8$ and $b=7a_{[110]}$.
For best illustration, we show diagonal strains  
in a rotated frame, in which $x'$ is [1$\bar 1$0], $y'$ is [110], 
and  $z'$ is [001].  The layer number in the figure labels the atomic
layers in Fig. 1(a), starting from the bottom layer as layer 1.
 There are two main features of the microscopic strain
distribution. First, 
in the ideal situation (without atomic relaxation), one would predict 
$\epsilon_{y'y'}$ to be the same as $\epsilon_{x'x'}$ due to symmetry. 
However, with atomic relaxation, all Ga(In) atoms tend to shift
in a direction so as to reduce the strain in the Ga(In)-rich region.
Thus, the magnitude of $\epsilon_{y'y'}$ on Ga(In) sites
(dashed lines) is lower(higher) than $\epsilon_{x'x'}$ in
Ga(In)-rich region.
Second, the $z$ component strain (dash-dotted lines) tend to 
compensate the other two components such that the volume of each bulk unit cell
is closer to that for the unstrained bulk.
Thus, we see that $\epsilon_{z'z'}$  has an opposite sign
compared to $\epsilon_{x'x'}$ or $\epsilon_{y'y'}$ at all atomic sites. 

Comparing Fig. 2 with Fig. 3, we see that the main difference between
the clamped and unclamped case is that in the atomic layers with
lateral composition modulation (layers 2 and 4), the hydrostatic strain
(sum of all three diagonal strain components) is much smaller in the 
unclamped structure than in the clamped structure. This would lead
to a major difference in the band-edge profile for the two structures
to be discussed below.

	The difference in $\epsilon_{x'x'}$ and $\epsilon_{y'y'}$ 
leads to a nonzero shear strain $e_{xy}=(\epsilon_{y'y'}-\epsilon_{x'x'})/2$
in the original coordinates. For the clamped case, 
the shear strain is found to be particularly strong near the boundary where
the alloy composition begins to change, 
and it is sensitive to the degree of lateral alloy mixing.
For $x_m=0.8$ and $b=7a_{[110]}$, the maximum value of
$\epsilon_{xy}$ is around 0.4\% 
For abrupt composition modulation ($x_m=1$ and $b=0$)
the  maximum $\epsilon_{xy}$ value increases five-fold to around 2\%.
The other shear strain components ($\epsilon_{xz}$ and
$\epsilon_{yz}$) are found to have similar magnitude.
For the unclamped case, the shear strain is strong even in regions
away from the boundary, and the maximum shear strain is larger than 
its counterpart in the clamped structure by about 30\%.

The diagonal strains of structure 2 for the unclamped case 
with $x_m=1$ and $b=7a_{[110]}$
are shown in Fig. 4. The layer number in the figure labels the atomic
layers in Fig. 1(b), starting from the bottom layer as layer 1.
Only four representative atomic layers (3,4,5, and 7)
are shown. We note that the average magnitude of the hydrostatic strain in the 
In-rich region in model structure 2 is comparable to that for 
unclamped structure 1.

{\bf B. Electronic structures}

    In order to understand the aspect of lateral quantum confinement
due to composition modulation and strain, we examine
the band-edge energies of a strained quantum well structure whose
well material is the same structure as appeared in the SILO
QWR with a fixed value of $y'$ and the barrier is 60\AA thick
Al$_{0.24}$Ga$_{0.24}$In$_{0.52}$As layer.
For structure 1 depicted in
Fig. 1(a), the well material consists of 8 pairs of
(GaAs)$_2$/(InAs)$_2$ SPS.
The conduction band minimum and valence band maximum of the above quantum well 
with different degrees of lateral alloy mixing
as functions of $y'$ (dashed lines) are shown in Fig. 5 for both clamped
and unclamped cases. 
The strain Hamiltonian used here is the same as the one used in
the self-assembled QWR at the corresponding $y'$.
All material parameters are chosen
the same as in Ref. [9] for temperature at 77K, except that the deformation
potential $c_1$ of GaAs is slightly modified from $-6.8eV$ to $-7.1eV$ so
that $a_1+a_2+c_1=-9.8eV$ is in agreement with the experimental measurements[26]
and the valence-band offset between GaAS and InAs used here is 0.26 eV 
according to the model solid theory[27].
To correct the band gap of SPS and alloy due to the bowing effect, we
add a correction term $-0.4x(1-x) eV$ to the diagonal element for the
s-like bond orbital in the Hamiltonian, where $x$ is the effective alloy
composition of the SPS or alloy at a given $y'$.
With the correction of bowing effect, our model gives band gaps for
the (GaAs)$_2$(InAs)$_2$ SPS, Ga$_{1-x}$In$_{x}$As alloy, and 
the superlattice made of (GaAs)$_2$(InAs)$_2$ SPS grown on InP substrate
all in very good agreement with available experimental observations.
The photoluminescence (PL) measurements
indicates that the (GaAs)$_2$(InAs)$_2$ SPS grown on InP
substrate has a gap around 0.75 eV.[28] Here we obtain a band gap of
0.74 eV for the (GaAs)$_2$(InAs)$_2$ SPS (with a strain distribution 
obtained again via the VFF model) and
0.79 eV for 8 pairs (or 100 \AA) of (GaAs)$_2$(InAs)$_2$ SPS 
sandwiched between Al$_{0.24}$Ga$_{0.24}$In$_{0.52}$As confining barriers.
This is also consistent with the PL measurements on the
(GaAs)$_2$(InAs)$_2$/InP multiple quantum wells.[28] 

The band-edge profiles shown in this figure suggest that for the clamped 
case with alloy mixing, both electrons
and holes are confined in the Ga-rich region with an effective barrier height
around $0.13 eV$ for the electron and $0.1 eV$ for the hole. 
Both offsets are large enough
to give rise to strong lateral confinement for electrons and holes in
the Ga-rich region. Without alloy mixing, the electron is not well confined,
since the average energy between the Ga-rich and In-rich regions are nearly
the same. For the unclamped case, both electrons
and holes are confined in the In-rich region with an effective barrier height
around $0.2 - 0.3 eV$ for the electron and $0.1 - 0.2 eV$ for the hole. 
We note that the band gap in the In-rich region is rather insensitive to
the alloy mixing, while the effective barrier height can change substantially
due to different alloy mixing.

Fig. 6 shows the band-edge profiles for 8 pairs of (2/2.25) SPS 
[as illustrated in Fig. 1(b)]
sandwiched between 60\AA Al$_{0.24}$Ga$_{0.24}$In$_{0.52}$As barriers
with and without alloy mixing for both clamped and unclamped cases.
For the clamped case, the electron is confined in the
In-rich region, while the hole is confined in the Ga-rich region. Thus, we 
have a spatially indirect QWR, which will have very weak inter-band optical 
transition.  For the unclamped case, the band-edge profile is similar to
the unclamped case of structure 1 with both the electron and hole confined in
the In-rich region. The band gaps for all cases of structure 2 are consistently
smaller than their counterparts in structure 1.    

Fig. 7 shows the near zone-center valence subband structures and squared 
optical matrix elements ($P^2$) for
the QWR model structure 1 [as depicted in Fig. 1(a)] with thick capping
layer (clamped case) and without alloy mixing.
Here $P^2$ is defined
as $\frac 2 {m_0} 
\sum_{s,s'} |<\psi_{v}|\hat \epsilon \cdot {\bf p}|\psi_{c}>|^2$,
where $\psi_{v}$ ($\psi_{c}$) denotes a valance (conduction) subband state.
The symbol $\sum_{s,s'}$ means a sum over two near degenerate pair of
subbands in the initial and final states. $\hat \epsilon$ denotes
the polarization of light. Here we consider only the $x'$ (along the
QWR axis) and $y'$ (perpendicular to the QWR axis) components.
Only the dispersion along the $k_1$ ($[1\bar 10]$) 
direction is shown, since
the dispersion along the $k_2$ direction for the confined levels is 
is rather small due to strong lateral confinement.
All subbands are two-fold degenerate at the zone center
due to the Kramer's degeneracy and
they split at finite wave vectors as a result of lack of
inversion symmetry in the system.
The first three pairs of subbands are labeled V1, V2, and V3.
They have unusually large energy separations compared with other
valence subbands. This is because the first three pairs of subbands represent
QWR confined states, whereas the other valence subbands (with energies below
-110 meV) are unconfined by the lateral composition modulation.

To examine the effects caused by the shear strain, we have also calculated
the band structures with the shear strain set to zero. We found that
without the shear strain the first pairs of valence subbands (V1) have 
much larger dispersion along the $k_1$ direction with an effective masses
along $[1\bar 10]$ about a factor five smaller than that  
with the shear strain. This indicates that the
V1 subband states for the case without shear strain are derived
mainly from bond orbitals with $x'$-character, which leads to stronger
overlap between two bond orbitals along the $x'$ direction, hence 
larger dispersion along $k_1$. When the shear strain is present, the character
of bond orbitals in the V1 subband states change from mainly $x'$-like 
to predominantly $y'$-like; thus, the dispersion along $k_1$ becomes
much weaker. This explains the very flat V1 subbands as shown in Fig. 6.
The switching of orbital character in the V1 subbands can be
understood as follows. When the  shear strain is absent, we have 
$\epsilon_{x'x'}=\epsilon_{y'y'}$, and the confinement effect in the $y'$
direction pushes down the states with $y'$ character (which has smaller
effective mass in the $y'$ direction), thus leaving the top valence 
subband (V1) to have predominantly $x'$ character. On the other hand,
with the presence of shear strain as shown in structure 1, 
we have $\epsilon_{y'y'} < \epsilon_{x'x'}$ in Ga-rich region, 
and the bi-axial term of the strain potential forces the $y'$-like states 
to move above the $x'$-like states, overcoming the confinement effect. 
As a result, the V1 subbands become $y'$-like.
The conduction subbands are approximately
parabolic as usual with a zone-center subband minimum equal to
688 meV 
This gives an energy gap of 763 meV  for QWR model structure 1.
As seen in this figure, the C1-V1 transition for $y'$-polarization (dashed
line) is more than twice that for the $x'$-polarization (solid line) 
with $P^2_{[1\bar 1  0]}/P^2_{[110]} \approx 0.36$.
This is consistent with the discussion for valence subband structures, 
where we concluded that the bond orbitals involved in the
V1 subbands are predominantly $y'$-like.

Next, we study the case with 
alloy mixing, in which a gradual lateral modulation in In composition 
substantially reduces the shear strain. Using a lateral alloy mixing
described in Eq. (1) with $x_m=0.8$ and $b=7a_{[110]}$,
we found significant change in band structures and optical properties
compared to the case without alloy mixing.
With alloy mixing we found that the first pairs of valence subbands 
(V1) have  much larger dispersion than their counterparts in Fig. 6, and
the  atomic character in the $V1$ states change from predominantly
$y'$-like (without alloy mixing) to $x'$-like (with alloy mixing).  
The change of atomic character in the V1 states
is a consequence of the competition between the QWR confinement (which
suppresses
the $y'$ character) and the $e_{xy}$ shear strain (which suppresses the $x'$
character). As a result, the ratio $P^2_{[1\bar 1  0]}/P^2_{[110]}$ changes from 0.36 
(without alloy mixing) to 3.5 (with alloy mixing), indicating a reversal 
of optical anisotropy.
The above studies lead to the conclusion that
for self-assembled QWRs with thick capping layer (clamped case), the optical
anisotropy is sensitive to the degree of lateral alloy mixing. 

Next, we study the properties of self-assembled QWRs with thin capping layer
(unclamped case). We found that for the unclamped case, the electron and hole are always confined in the In-rich region, regardless of the degree of alloy mixing and whether the QWR structure consists of (2/2) SPS or (2/2.25) SPS. Furthermore, since the strain in In-rich region has opposite sign compared to
that in the Ga-rich region, the effect of shear strain on the VB states confined in In-rich region is also reversed. Namely, it pushes the energy of the
$y'$-like VB states down relative to the $x'$-like VB states. Thus, both the QWR
confinement and the shear strain effects are in favor of having a predominant 
$x'$ character in the V1 subband. Therefore, the inter-band optical 
transition always has a ratio $P^2_{[1\bar 1  0]}/P^2_{[110]}$ larger than 1.
  
To illustrate this, we show in Fig. 8 the near zone-center valence subband structures and squared optical matrix elements ($P^2$) for
the QWR model structure 1 [as depicted in Fig. 1(a)] with thin capping
layer (unclamped case) and with alloy mixing described by $x_m=0.8$ and
$b=7a_{[110]}$. 
Comparing the band structure in Fig. 8 with that in Fig. 7, we
see that the V1 subband now has much larger dispersion (i.e. much smaller
effective mass) than its counterpart in Fig. 6. There are two reasons for this.
First, the hole in the unclamped case is now confined in the In-rich region
rather than in the Ga-rich region as in the clamped case, thus having
much smaller effective mass. Second, both the QWR confinement
effect and the shear strain effect lead to a predominantly $x'$ character in the V1 states, which tend to yield a smaller effective mass in the $k_1$ direction.
The optical properties shown in Fig. 8 are also distinctly different from that 
shown in Fig. 7. The inter-band optical transition (C1-V1) now has a
much stronger polarization along [1$\bar 10$] than [110]. 
The quick drop of the the $y'$ polarization in C1-V1 transition
at $k_1\approx 0.035 \frac {2\pi} a$ is caused by the band
mixing of V1 and V2 states
as can be seen in the VB band structures in this figure.

We have also calculated the near zone-center valence subband structures and squared 
optical matrix elements ($P^2$) for
the QWR model structure 2 with thin capping
layer (not shown). We found that the band structures and optical matrix elements
are similar to Fig. 8, except that the energy separation between V1 and V2 subbands are smaller (by $\sim 5 mV$) and the polarization ratio ($P^2_{[1\bar 1  0]}/P^2_{[110]}$) is larger. 

	Finally, we list in Table I
the band gaps and the polarization ratio ($P^2_{[1\bar 1  0]}/P^2_{[110]}$) 
for various
QWR structures with different degrees of lateral alloy mixing (as indicated by
the parameters $x_m$ and $b$).
To calculate the polarization ratio in the photoluminescence (PL) spectra, we integrate the squared optical matrix
element over the range of $k_1$ corresponding to the spread of exciton 
envelope function in the $k_1$ space. The exciton envelope function
is obtained by solving the 1D Schr\"{o}dinger equation for the exciton in
the effective-mass approximation similar to what we did in Ref. 9.
For unclamped (2/2) SPS structure, the band gap varies between 0.77 and 0.8 eV and the polarization ratio changes from 3.1 to 1.5 as the degree of alloy mixing is increased. For unclamped (2/2.25) SPS structure, both the band gap (around 0.74 eV) and the 
polarization ratio (around 2) are insensitive to the degree of alloy mixing.

	Experimentally, the PL peak for self-assembled InGaAs QWRs
with similar dimensions as considered in our model structures
are around 735 meV and the polarization ratio for different samples are 
found to be between 2 and 4.[4-6] Taking into account an exciton binding 
energy between 10 and 20 meV, we find that our theoretical predictions 
for both unclamped and clamped structures with lateral alloy mixing are 
consistent with the experimental finding. To sort out which one is closer 
to the experimental structure would require more detailed comparison 
between the theory and experiment for optical spectra involving the 
excited hole states such as the PL excitation spectra.  
 
\section{Conclusion}

We have calculated the band structures and optical matrix elements 
for the self-assembled GaInAs  QWR grown by the SILO method. The actual SPS
structure and the microscopic strain distribution has been
taken into account. The effects of microscopic strain  distribution
on the valence subband structures and optical matrix elements are studied
for two model self-assembled QWR structures, one with (2/2) SPS structure and the other with (2/2.25) SPS structure, and with varying degrees of lateral alloy mixing. The clamping effect due to thick capping layer is also examined. 
The valence force field (VFF) model is used to calculate the equilibrium
atomic positions in the model QWR structures. This allows the calculation
of the strain distribution at the atomistic level.

We found that in model structures with thick capping layer (clamped case),
the magnitude of shear strain is quite large (around 2\%) for abrupt lateral
composition modulation and it reduces substantially when the lateral alloy mixing is introduced.  We found that the shear strain effect can 
alter the valence subband structures substantially when the hole is confined in the Ga-rich region, and it gives rise
to a reversed optical anisotropy compared with QWR structures with
negligible shear strain. 
This points to a possibility of "shear strain-engineering" to obtain
QWR laser structures of desired optical anisotropy.
For model structures with thin capping layer (unclamped case), the effect of
shear strain is less dramatic, since the hole is confined in the In-rich region,
where both the QWR confinement and the shear strain have similar effect. We found that polarization ratio ($P^2_{[1\bar 1  0]}/P^2_{[110]}$) for the unclamped case is always larger than 1 for both (2/2) and (2/2.25) SPS structures, regardless of the degree of lateral alloy mixing. 

    The band gap and polarization ratio obtained for both clamped and unclamped (2/2) SPS structures with lateral alloy mixing ($x_m \approx 0.8$) and for unclamped (2/2.25) SPS structures  are consistent
with experimental observations on most self-assembled InGaAs 
QWRs with similar specifications. For clamped (2/2) SPS structure without alloy mixing, we predicted a reversed optical anisotropy compared with the PL
measurements for most InAs/GaAs self-assembled QWRs. However, there exist a few
InAs/GaAs self-assembled QWR samples which display the reversed optical 
anisotropy at 77 K[7], and recent experimental studies also indicated a reversed
optical anisotropy in many samples at 300 K.[6] Our studies demonstrated 
one possibility for the reversed optical anisotropy, the shear strain effect. 
There are other possibilities for the reversed optical anisotropy at 300K, such as the
thermal population of the excited hole states, which can have different 
atomic characters as discussed in the previous section. 
The temperature dependence of the PL peak and polarization in 
self-assembled InGaAs QWRs[6] remains an intriguing question to be 
addressed in the future.

\vspace{1cm}
\leftline{\bf ACKNOWLEDGEMENTS}
\vspace{1ex}
  This work was supported in part by the National Science Foundation (NSF) 
under Grant No. NSF-ECS96-17153. S. Sun was also supported by
a subcontract from the University of Southern California under the
MURI program, AFOSR, Contract  No. F49620-98-1-0474.
We would like to thank K. Y. Cheng
and D. E. Wohlert for fruitful discussions
and for providing us with the detailed experimental data
of the QWR structures considered here. 

\newpage

{\bf Table I. \hspace{0.2in} List of band gaps and polarization ratios for 
all structures studied}

\vspace{1ex}
\begin{tabular}{lllllll} \hline\hline
$x_m$ \hspace{0.4in} & b/$a_{110}$\hspace{0.2in} & alloy mixing
\hspace{0.2in} & clamped/unclamped\hspace{0.2in} & SPS structure\hspace{0.2in} & gap(eV)\hspace{0.2in} & $P_{\parallel}/P_{\perp}$ \\ \hline
 1.0  & 0   & no       & clamped       & (2/2)         & 0.765   & 0.36  \\ 
0.8   & 7  & yes       & clamped        & (2/2)         & 0.763   & 3.5  \\ 
 1.0  & 0   & no       & unclamped     & (2/2)         & 0.791   & 3.1  \\ 
 0.7  & 7  & yes       & unclamped      & (2/2)         & 0.766   & 1.8  \\ 
 0.6  & 7  & yes       & unclamped      & (2/2)         & 0.80    & 1.82 \\ 
0.6   & 15 & yes       & unclamped      & (2/2)         & 0.798   & 1.5  \\ 
1.0   & 0   & no       & unclamped     & (2/2.25)      & 0.731   & 2.2 \\
1.0   & 7  & yes       & unclamped      & (2/2.25)       & 0.74    & 2.1 \\ 
\hline\hline
\end{tabular}

\newpage

\vspace{2cm}
Figure Captions

\vspace{1ex}

\noindent Fig. 1.  Schematic sketch of the unit cell of
the SILO quantum wire for two model structures considered. 
Each unit cell consists of 8 pairs of (2/2) or (2/2.25) GaAs/InAs 
short-period superlattices (SPS).
In structure 1, the (2/2) SPS contains four diatomic layers (only the cation
planes are illustrated) and the SPS is repeated eight times. 
In structure 2, four pairs of (2/2.25) SPS (or 17 diatomic layers) form
a period, and the period is repeated twice in the unit cell.  
Filled and open circles indicate Ga and In rows 
(each row extends infinitely along the $[1\bar 1 0]$ direction). 

\noindent Fig. 2. Diagonal strain distribution of model structure 1
[corresponding to Fig. 1(a)] with thick capping layer (clamped case).
Solid: $\epsilon_{x'x'}$. Dashed: $\epsilon_{y'y'}$.
Dash-dotted: $\epsilon_{zz}$. Parameters for alloy mixing: $x_m=0.8$ and 
$b=7 a_{[110]}$.

\noindent Fig. 3. Diagonal strain distribution of model structure 1
with thin capping layer (unclamped case).
Solid: $\epsilon_{x'x'}$. Dashed: $\epsilon_{y'y'}$.
Dash-dotted: $\epsilon_{z'z'}$.
Parameters for alloy mixing: $x_m=0.8$ and $b=7 a_{[110]}$.
         
\noindent Fig. 4. Diagonal strain distribution of model structure 2
[corresponding to Fig. 1(b)] with thin capping layer (unclamped case).
Solid: $\epsilon_{x'x'}$. Dashed: $\epsilon_{y'y'}$.
Dash-dotted: $\epsilon_{z'z'}$.
Parameters for alloy mixing: $x_m=0.8$ and $b=7 a_{[110]}$.

\noindent Fig. 5. Conduction and valence band edges
for 8 pairs of (2/2) SPS structures
sandwiched between 60\AA Al$_{0.24}$Ga$_{0.24}$In$_{0.52}$As barriers
with different degrees of lateral 
alloy mixing as functions of the [110] coordinate $y'$ for both
clamped (thick capping layer) and unclamped (thin capping layer) case.
In upper pane, the dashed line is for abrupt modulation (without alloy mixing)
and the solid line is for lateral alloy mixing with $x_m=0.8$ and $b=7 a_{[110]}$.
In lower panel, dash-dotted line: without alloy mixing; solid line:
$x_m=0.7$ and $b=7 a_{[110]}$;
dotted lines: $x_m=0.6$ and $b=7 a_{[110]}$;
dashed lines: $x_m=0.6$ and $b=15 a_{[110]}$.

\noindent Fig. 6. Conduction and valence band edges
for 8 pairs of (2/2.25) SPS structures
sandwiched between 60\AA Al$_{0.24}$Ga$_{0.24}$In$_{0.52}$As barriers
with different degrees of lateral 
alloy mixing as functions of the [110] coordinate $y'$ for both
clamped (thick capping layer) and unclamped (thin capping layer) case.
Dashed lines: without alloy mixing.
Solid lines: with alloy mixing described by $x_m=1.0$ and $b=7 a_{[110]}$.

\noindent Fig. 7. Valance subband structures and squared optical matrix elements
for self-assembled QWR made of (2/2) SPS structure                
with thick capping layer (clamped case) and without alloy mixing.      

\noindent Fig. 8. Valance subband structures and squared optical matrix elements
for self-assembled QWR made of (2/2) SPS structure              
with thin capping layer (unclamped case) and with alloy mixing described
by $x_m=0.8$ and $b=7 a_{[110]}$.


\newpage

\begin{thebibliography}{wide label}

\bibitem{adams}A. R. Adams, Electron. Lett. {\bf 22}, 249 (1986).
\bibitem{yablonocitch}E. Yablonovitch and E. O. Kane. IEEE J. Lightwave 
Technol. LT-4, 504 (1986).
\bibitem{agrawa}G. P. Agrawa and N. K. Dutta, Long Wavelength Semiconductor
 Lasers, 2nd ed. (Van. Nostrand Reinhold, New York, 1993) Chap.7.
\bibitem{chou}S.T. Chou, K. Y. Cheng, L. J. Chou, and K. C. Hsieh, Appl. 
Phys. Lett. {\bf 17}, 2220 (1995); J. Appl. Phys. {\bf 78} 6270, (1995); 
J. Vac. Sci. Tech. B {\bf 13}, 650 (1995);
K. Y. Cheng, K. C. Hsien, and J. N. Baillargeon, Appl. Phys. Lett. {\bf 60}, 
2892 (1992).
\bibitem{wohlert}D. E. Wohlert, S. T. Chou, A. C Chen, K. Y. Cheng, 
and K. C. Hsieh, Appl. Phys. Lett. {\bf 17}, 2386 (1996).
\bibitem{wohlert1} D.E. Wohlert, and K. Y. Cheng, Appl. Phys. Lett. {\bf 76}, 2249 (2000).
\bibitem{wohlert2} D.E. Wohlert, and K. Y. Cheng, private communications. 
\bibitem{rich} Y. Tang, H. T. Lin, D. H. Rich, P. Colter, and S. M. Vernon,
Phys. Rev. B {\bf 53}, R10501 (1996).
\bibitem{masc} Y. Zhang and A. Mascarenhas, Phys. Rev. B{\bf 57}, 12245
(1998).
\bibitem{li}L. X. Li and Y. C. Chang, J. Appl. Phys. {\bf 84} 6162, 1998. 
\bibitem{Zun1} T. Mattila, L. Bellaiche, L. W. Wang, and A. Zunger, Appl.
Phys. Lett., {\bf 72}, 2144 (1998).
\bibitem{Zun2} L. W. Wang, J. Kim, and A. Zunger, Phys. Rev. B {\bf 59}, 5678
(1999).
\bibitem{jiang}H. Jiang and J. Singh, Phys. Rev. B{\bf 56}, 4696(1997).
\bibitem{keating}P. N. Keating, Phys. Rev. {\bf 145}, 637(1966).
\bibitem{martin}R. M. Martin, Phys. Rev. B{\bf 1}, 4005(1969).
\bibitem{chang}Y. C. Chang, Phys. Rev. B{\bf 37}, 8215 (1988).
\bibitem{houng}Mau-Phon Houng and Y.C Chang, J. Appl. Phys. {\bf 65}, 3096
(1989).
\bibitem {Gru} M. Grundmann, O.Stier, and D. Bimberg, 
Phys. Rev. B {\bf 52},11969 (1995).
\bibitem {Sti} O.Stier, M.Grundmann, and D.Bimberg, 
Phys. Rev. B {\bf 59}, 5688,(1999).
\bibitem{bir}G. L. Bir and G. E. Pikus, Symmetry and Strain Induced Effects 
in Semiconductors (Halsted, United Kingdom, 1974); 
L.D. Landau and E. L. Lifshitz, Theory of Elasticity 
(Addison-Wesley Publishing Company Inc, Reading, Massachusetts, USA, 1970).
\bibitem{pryor}C. Pryor, J. Kim, L.W. Wang, A.J. Williamson and A. Zunger, Phys. Rev. B{\bf 183}, 2548(1998).
\bibitem{adachi}S. Adachi, J. Appl. Phys. {\bf 53}, 8775 (1982).
\bibitem{mathieu}H. Mathieu, P. Meroe, E. L. Amerziane, B. Archilla, J. 
Camassel, and G. Poiblaud, Phys. Rev. B{\bf 19}, 2209 (1979); 
S. Adachi and C. Hamaguchi, Phys. Rev. B{\bf 19}, 938 (1979). 
\bibitem{hinckley}J. M. Hinckley and J. Singh, Phys. Rev. B{\bf 42},
3546 (1990).
\bibitem{land}O. Madelung and M. Schulz, Landolt-B\"{o}rstein, New Series
III/22a, p. 87 (1982).
\bibitem{Cardona} A. Blacha, H. Presting, M Cardona, Phys. Stat. Sol. (b) {\bf
126}, 11 (1984).
\bibitem{Walle} C. G. Van de Walle, Phys. Rev. B {\bf 39}, 1871 (1989).
\bibitem{Eaz} M. Razeghi, Ph. Maurel, F. Omnes, and J. Nagle, Appl. Phys.
Lett. {\bf 51}, 2216 (1987).





\end{thebibliography}
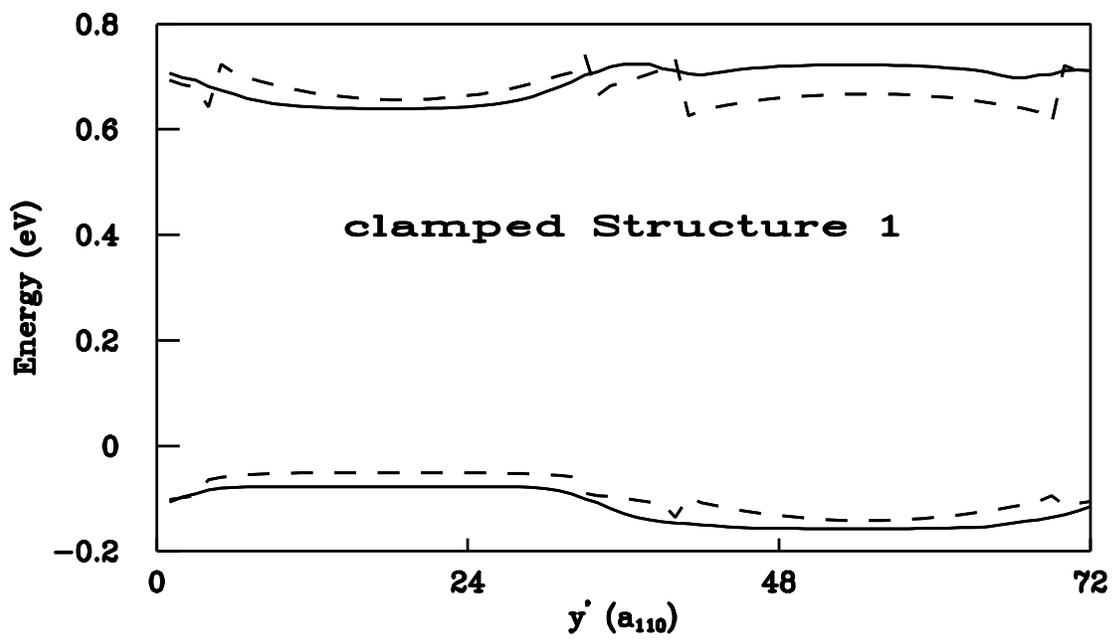
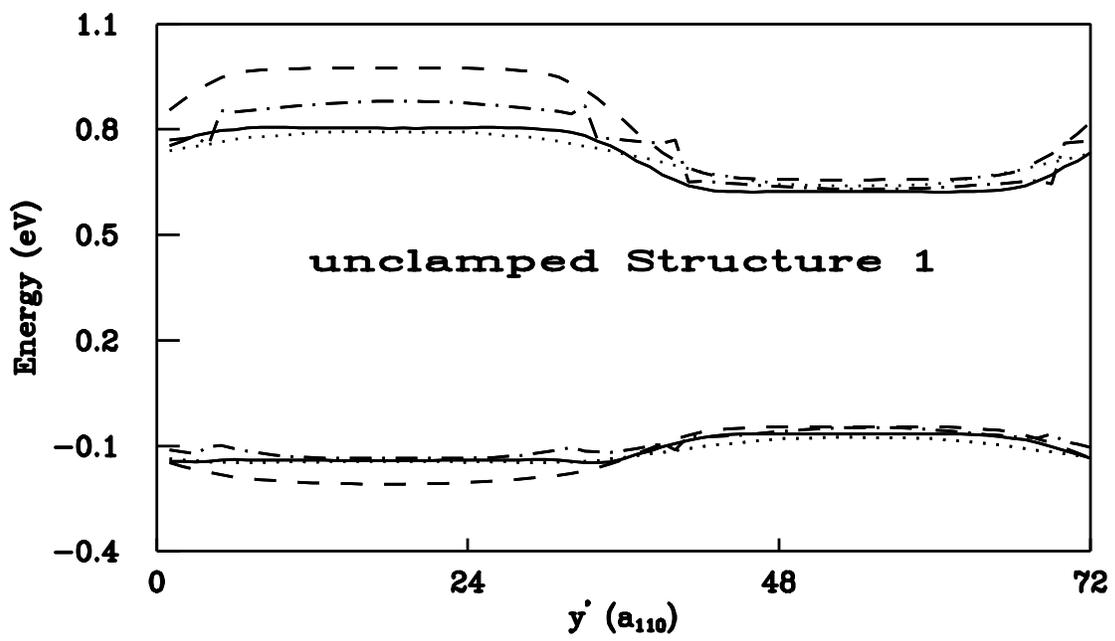
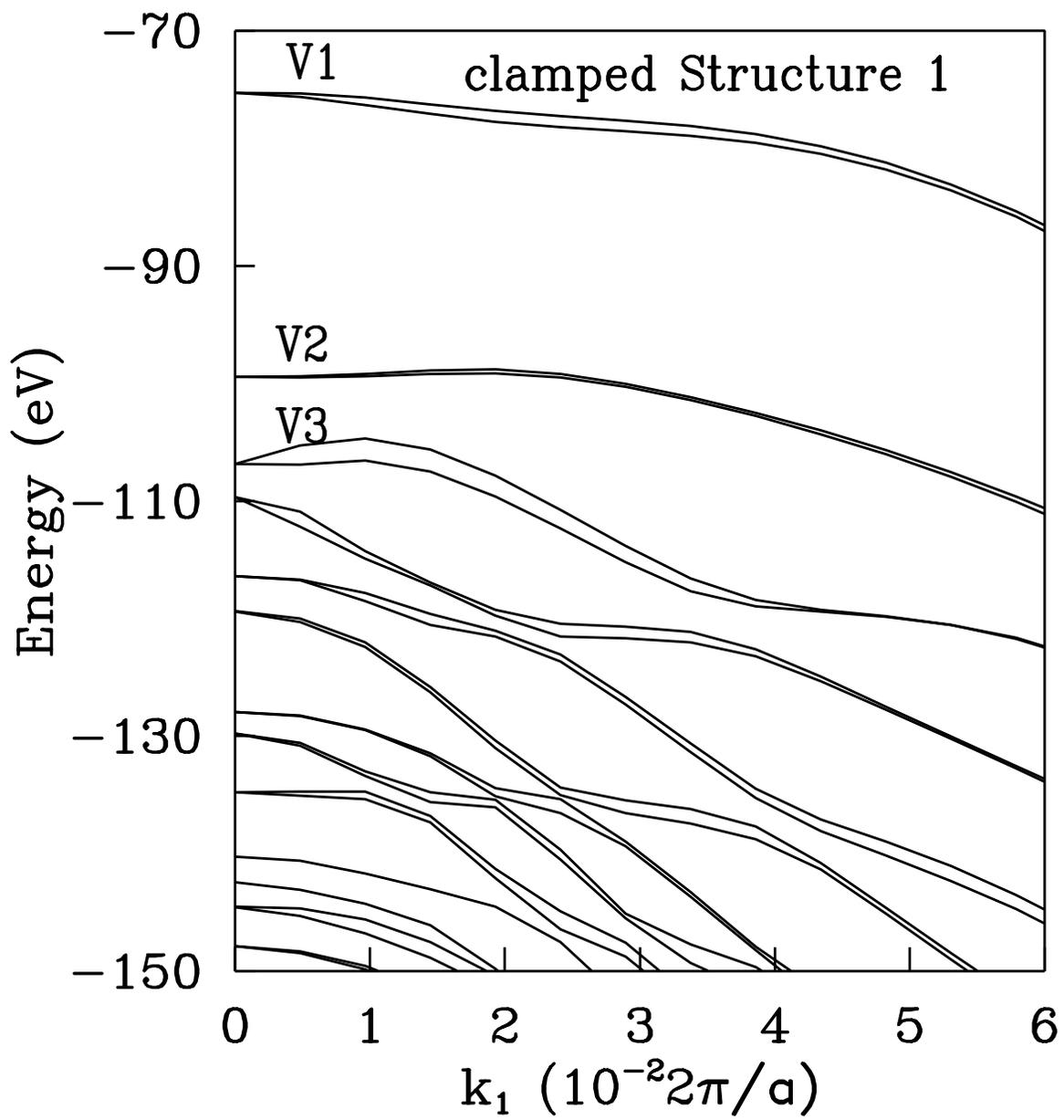
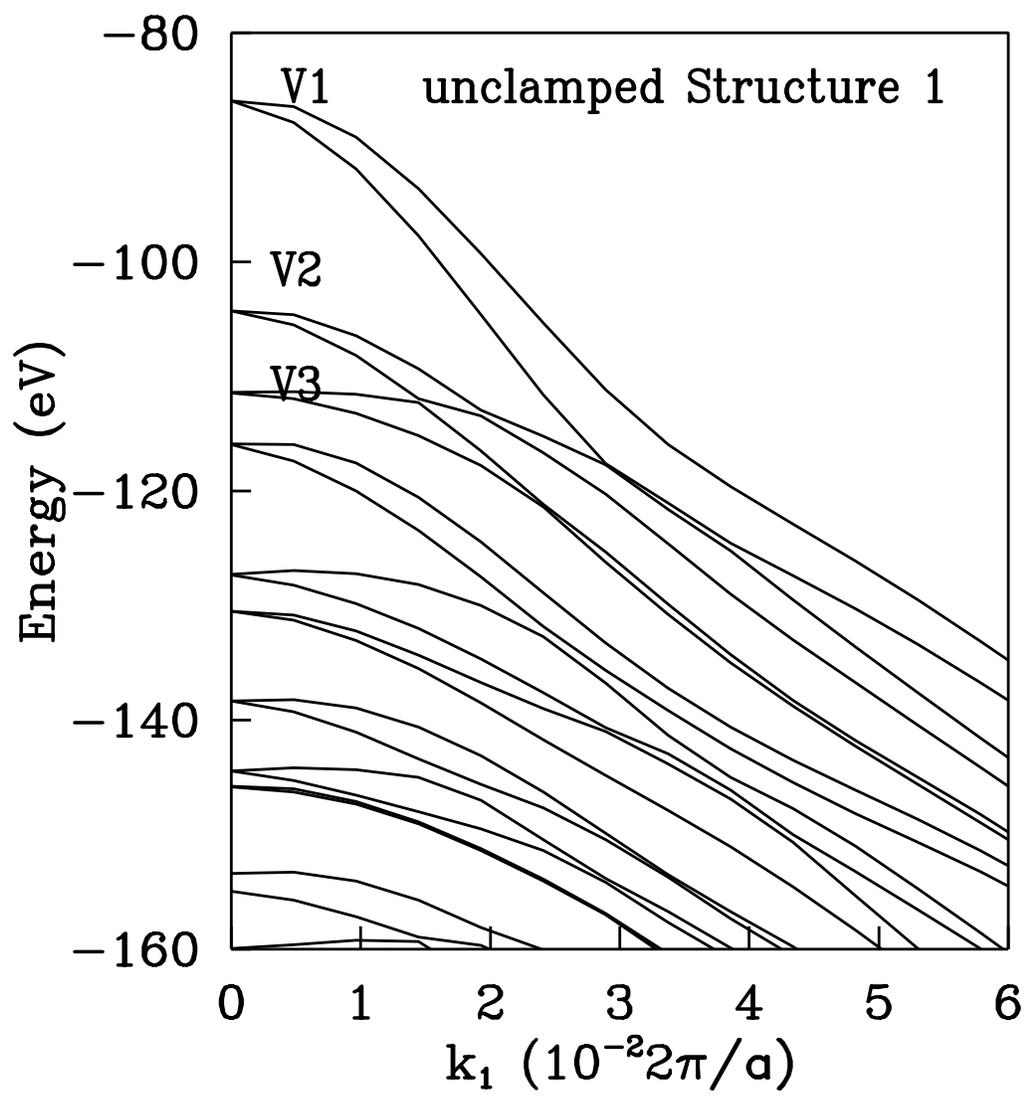
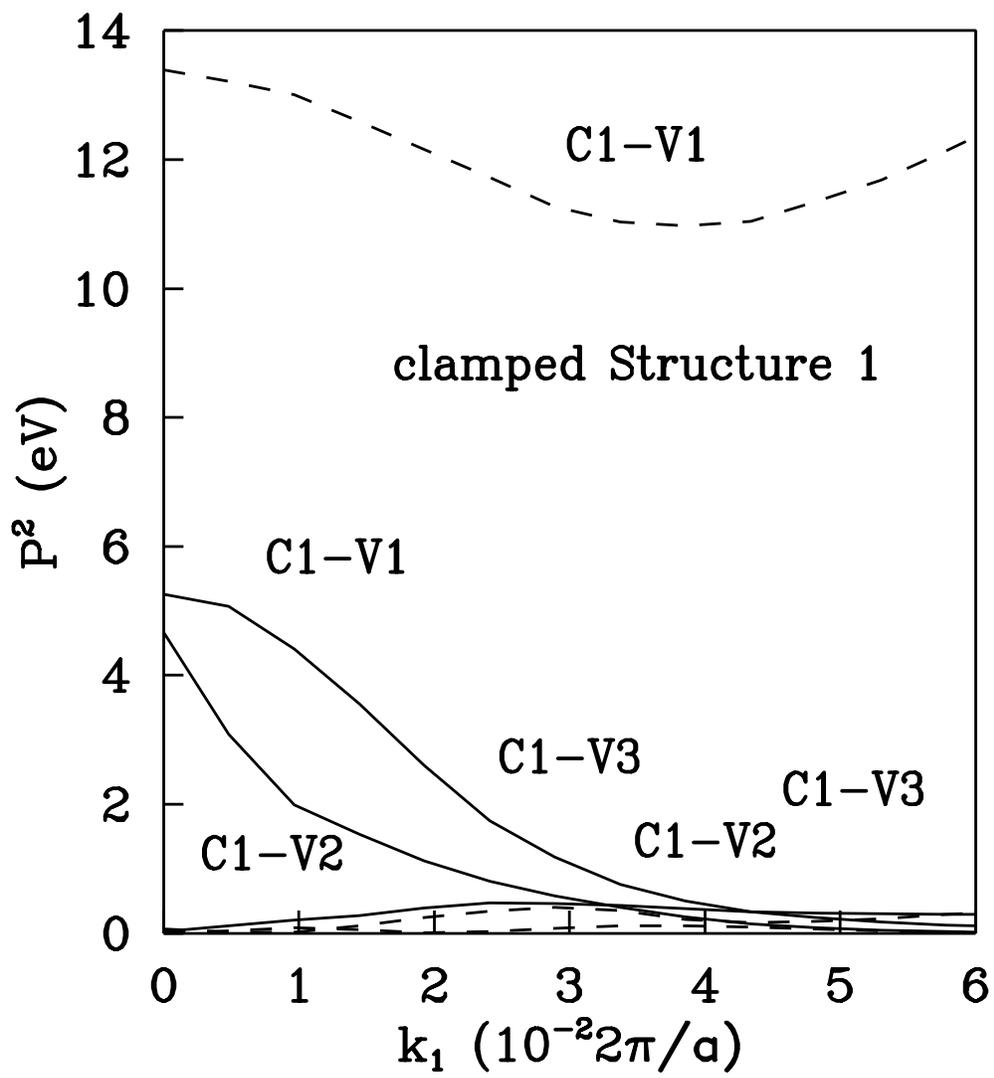
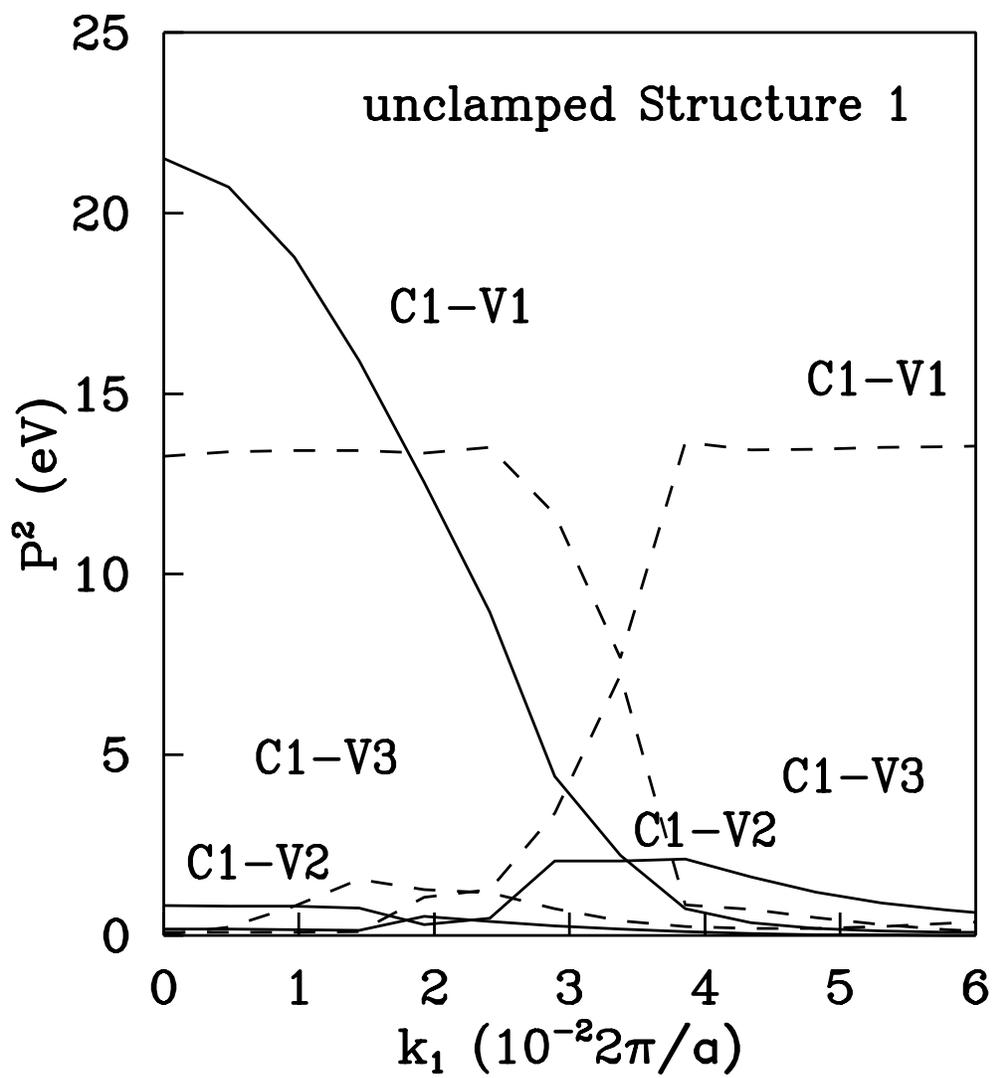
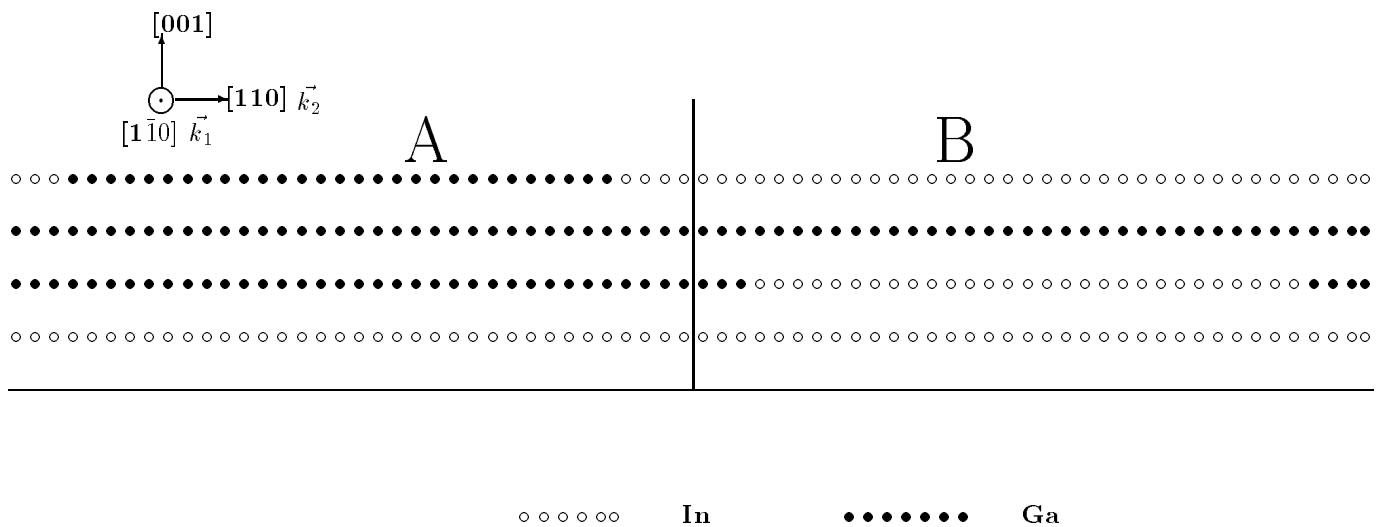
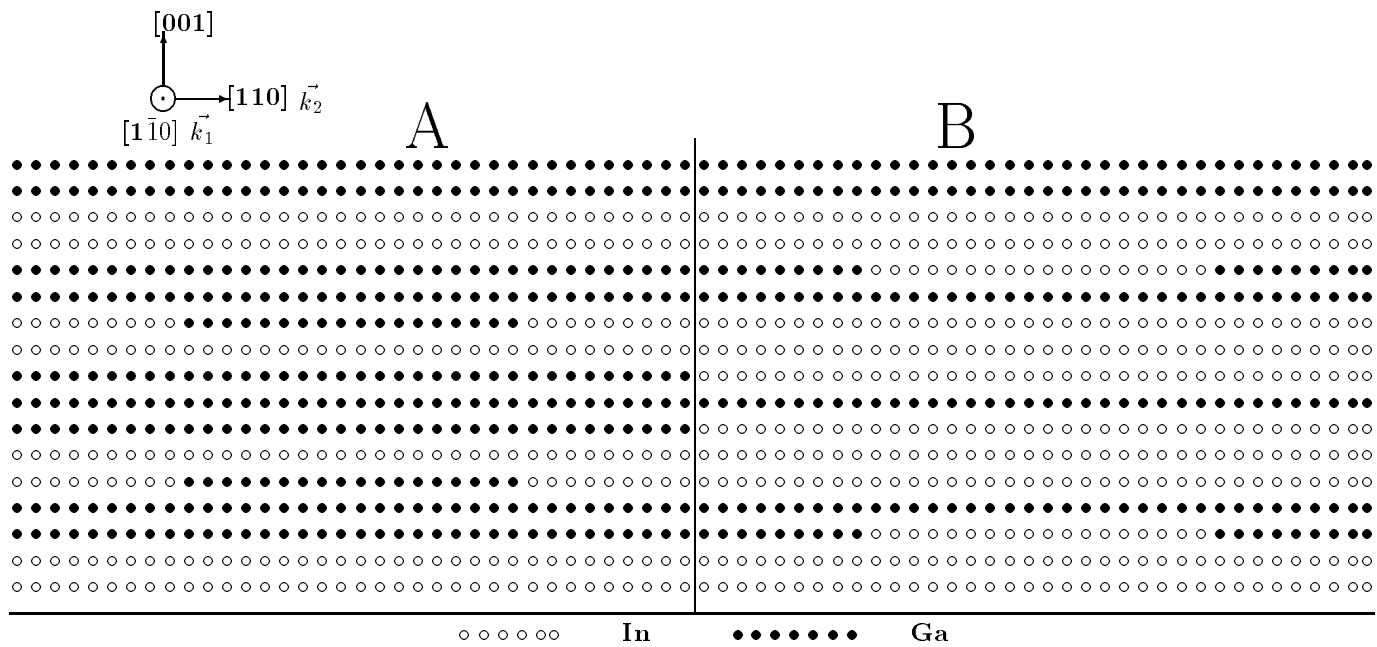
\end{document}